\documentclass[12pt]{article}
\usepackage{amssymb}
\renewcommand{\theequation}{\arabic{section}.\arabic{equation}}

\begin{document}



\def\a{\alpha}
\def\b{\beta}
\def\d{\delta}
\def\e{\epsilon}
\def\g{\gamma}
\def\h{\mathfrak{h}}
\def\k{\kappa}
\def\l{\lambda}
\def\o{\omega}
\def\p{\wp}
\def\r{\rho}
\def\t{\theta}
\def\s{\sigma}
\def\z{\zeta}
\def\x{\xi}
 \def\A{{\cal{A}}}
 \def\B{{\cal{B}}}
 \def\C{{\cal{C}}}
 \def\D{{\cal{D}}}
\def\G{\Gamma}
\def\K{{\cal{K}}}
\def\O{\Omega}
\def\R{\bar{R}}
\def\T{{\cal{T}}}
\def\L{\Lambda}
\def\f{E_{\tau,\eta}(sl_2)}
\def\E{E_{\tau,\eta}(sl_n)}
\def\Zb{\mathbb{Z}}
\def\Cb{\mathbb{C}}

\def\R{\overline{R}}

\def\beq{\begin{equation}}
\def\eeq{\end{equation}}
\def\bea{\begin{eqnarray}}
\def\eea{\end{eqnarray}}
\def\ba{\begin{array}}
\def\ea{\end{array}}
\def\no{\nonumber}
\def\le{\langle}
\def\re{\rangle}
\def\lt{\left}
\def\rt{\right}

\newtheorem{Theorem}{Theorem}
\newtheorem{Definition}{Definition}
\newtheorem{Proposition}{Proposition}
\newtheorem{Lemma}{Lemma}
\newtheorem{Corollary}{Corollary}
\newcommand{\proof}[1]{{\bf Proof. }
        #1\begin{flushright}$\Box$\end{flushright}}

\baselineskip=20pt

\newfont{\elevenmib}{cmmib10 scaled\magstep1}
\newcommand{\preprint}{
   \begin{flushleft}
   \end{flushleft}\vspace{-1.3cm}
   \begin{flushright}\normalsize
   \end{flushright}}
\newcommand{\Title}[1]{{\baselineskip=26pt
   \begin{center} \Large \bf #1 \\ \ \\ \end{center}}}
\newcommand{\Author}{\begin{center}
   \large \bf
Wen-Li Yang ${}^{a,b}$, and
  Yao-Zhong Zhang ${}^{b}$ \end{center}}
\newcommand{\Address}{\begin{center}

${}^a$ Institute of Modern Physics, Northwest University,
       Xian 710069, P.R. China\\
${}^b$ Department of Mathematics, University of Queensland, Brisbane,
       QLD 4072, Australia\\

E-mail: \,wenli@maths.uq.edu.au,\,\, yzz@maths.uq.edu.au

   \end{center}}
\newcommand{\Accepted}[1]{\begin{center}
   {\large \sf #1}\\ \vspace{1mm}{\small \sf Accepted for Publication}
   \end{center}}

\preprint
\thispagestyle{empty}
\bigskip\bigskip\bigskip

\Title{On explicit free field realizations of current algebras}
\Author

\Address
\vspace{1cm}

\begin{abstract}
We construct the explicit free field representations of the
current algebras $so(2n)_k$, $so(2n+1)_k$ and $sp(2n)_k$ for a
generic positive integer $n$ and an arbitrary level $k$. The
corresponding energy-momentum tensors and screening currents of
the first kind are also given in terms of free fields.

\vspace{1truecm} \noindent {\it PACS:} 11.25.Hf

\end{abstract}
\newpage
\section{Introduction}
\label{intro} \setcounter{equation}{0}

Conformal field theories (CFTs) \cite{Bel84,Fra97} have played a
fundamental role in the framework of string theory and the
theories of modern condensed matter physics and statistical
physics at critical point. The Wess-Zumino-Novikov-Witten(WZNW)
models \cite{Fra97}, whose symmetry algebras are current algebras
\cite{Kac90}, stand out as an important class of CFTs because
these models are ``building blocks" of all rational CFTs through
the so-called GKO coset construction \cite{God86}. The Wakimoto
free field realizations of current algebras \cite{Wak86,Fra97}
have been proven to be  powerful in the study of the WZNW models
\cite{Dos84,Fat86,Ber90,Fur93,And95} due to the fact that an {\it
explicit} free field representation enables one to construct
integral representations of correlators of the CFT.

Free field realizations of  current algebras have been extensively
investigated by many authors
\cite{Fei90,Bou90,Ber89,Ito90,Ger90,Fre94,Boe97,Ras97}. However,
it is very complicated to apply the general procedure proposed in
these references to derive explicit free field expressions of the
affine currents for higher-rank algebras \cite{Ito90, Ras97,
Din03,Zha05, Ras98}. To our knowledge, explicit expressions have
so far been known only for some isolated cases: those associated
with Lie algebras $su(n)$ \cite{Fei90}, $B_2$ (or $so(5)$)
\cite{Bou90}, $G_2$ \cite{Ito90},  and Lie superalgebra $gl(m|n)$
\cite{Yan07}. In particular, explicit free field expressions for
affine currents associated with Lie algebras $so(2n)$, $so(2n+1)$
and $sp(2n)$ for generic $n$ are still lacking\footnote{The
authors in \cite{Kuw90} proposed certain explicit free field
expressions for the $so(2n)$, $so(2n+1)$ and $sp(2n)$ current
algebras, but one can check that their results are incorrect, as
was also pointed out in \cite{Ito90}.}.

In this paper, we find a way to overcome the complication in the
above-mentioned general method. In our approach, the construction
of the differential operator realization becomes much simpler
(c.f. \cite{Ito90,Ras97, Din03}). We shall work  out the explicit
forms of the differential realizations of $so(2n)$, $so(2n+1)$ and
$sp(2n)$, and apply them to construct explicit free field
representations of the corresponding current algebras. These
representations provide the Verma modules of the algebras.

This paper is organized as follows. In section 2, we briefly
review the definitions of finite-dimensional Lie algebras  and the
associated current algebras, which also introduces  our notation
and some basic ingredients. In section 3, after constructing
explicitly the differential operator realization of $so(2n)$,  we
construct the explicit free field representations of the $so(2n)$
currents, the corresponding energy-momentum tensor and the
associated screening currents at a generic level $k$. In sections
4 and 5, we present the corresponding results for the $so(2n+1)$
and $sp(2n)$ current algebras, respectively. Section 6 is for
conclusions. In the Appendices A-C, we give the matrix
realizations associated with the fundamental representation for
all generators of the finite dimensional Lie algebras $so(2n)$,
$so(2n+1)$ and $sp(2n)$.


\section{Notation and prelimilaries}
\label{CUR} \setcounter{equation}{0}

Let $\textbf{g}$ be a simple Lie algebra with a finite dimension
dim$(\textbf{g})=d<\infty$ and $\lt\{E_i|i=1,\ldots,d\rt\}$ be a
basis of $\textbf{g}$. The generators $\lt\{E_i\rt\}$ satisfy
commutation relations, \bea
 \lt[E_i,\,E_j\rt]=\sum_{m=1}^df_{ij}^m\,E_m,\label{construct-constant}
\eea where $f_{ij}^m$ are the structure constants of $\textbf{g}$.
Alternatively, one can  use the associated root system
\cite{Fra96} to label the generators of $\textbf{g}$ as follows.
Let us assume the rank of $\textbf{g}$ to be rank$(\textbf{g})=n$
with a generic positive integer $n\geq 1$, and $\textbf{h}$ be a
Cartan subalgebra of $\textbf{g}$. The set of (positive) roots is
denoted by ($\Delta_+$) $\Delta$, and we write $\a>0$ if
$\a\in\Delta_+$. Among the positive roots, the simple roots are
$\lt\{\a_i|i=1,\ldots,n\rt\}$. Associated with each positive root
$\a$, there are a raising operator $E_{\a}$, a lowering operator
$F_{\a}$ and a Cartan generator $H_{\a}$. Then one has the
Cartan-Weyl decomposition of $\textbf{g}$ \bea
 \textbf{g}=\textbf{g}_{-}\oplus\textbf{h}
 \oplus\textbf{g}_+.
\eea

One can introduce  a nondegenerate and invariant symmetric metric
or bilinear form $\lt(E_i,E_j\rt)$ for $\textbf{g}$. For $so(2n)$,
$so(2n+1)$ and $sp(2n)$, which we consider in this paper, the
corresponding bilinear forms are given in (\ref{Bilinear-D}),
(\ref{Bilinear-B}) and (\ref{Bilinear-C}) respectively. Then the
current algebra  $\textbf{g}_k$ is generated by the currents
$E_i(z)$ associated with the generators $E_i$ of $\textbf{g}$. The
current algebra at a general level $k$ obeys the following OPEs
\cite{Fra97}, \bea
 E_i(z)E_j(w)=k\frac{(E_i,E_j)}{(z-w)^2}
 +\frac{\sum_{m=1}^df_{ij}^mE_m(w)}{(z-w)},\qquad i,j=1,\ldots,d,
 \label{current-OPE}
\eea where $f_{ij}^m$ are the structure constants
(\ref{construct-constant}). The aim of this paper is to construct
explicit free field realizations of the current algebras
associated with $so(2n)$, $so(2n+1)$ and $sp(2n)$ at a generic
level $k$.


\section{Free field realization of the $so(2n)$ currents}
 \label{DIR} \setcounter{equation}{0}

As mentioned in the introduction, practically it would be very
involved (if not impossible) to obtain  {\it explicit} free field
realizations of current algebras associated with  higher-rank
algebras by the general method outlined in
\cite{Fei90,Bou90,Ito90,Fre94,Ras97}. We have found a way to
overcome the complication. In our approach, the construction of
differential operator realizations becomes much simpler, giving
rise to explicit expressions of differential operators (see
(\ref{Diff-D-1})-(\ref{Diff-D-2}) for $so(2n)$,
(\ref{Diff-B-1})-(\ref{Diff-B-2}) for $so(2n+1)$ and
(\ref{Diff-C-1})-(\ref{Diff-C-2}) for $sp(2n)$  below). In this
section, we consider the $so(2n)$ current algebra for generic $n$
and arbitrary level $k$.

\subsection{Differential operator realization of $so(2n)$}


The root system of $D_n$ (or $so(2n)$) are: $ \pm\e_i\pm \e_j$,
for $i\neq j$ and $i,j=1,\ldots,n$. Among them, the positive roots
$\Delta_+$ can be chosen as: $\e_i\pm\e_j$ for $1\leq i<j\leq n$.
The simple roots are  \bea
 \a_1=\e_1-\e_2,\,\a_2=\e_2-\e_3,\ldots,
 \a_{n-1}=\e_{n-1}-\e_n,\,\a_n=\e_{n-1}+\e_n.\label{simple-roots-D}
\eea Hereafter, we adopt the convention that \bea
 E_i\equiv E_{\a_i},\quad F_i\equiv F_{\a_i}, \qquad
 i=1,\ldots,n.\label{convent}
\eea The matrix realization of the generators associated with all
roots of $so(2n)$ is given in Appendix A, from which one may
derive the structure constants  for the particular choice of the
basis.

Let us introduce a coordinate $x_{i,j}$ associated with  each
positive root $\e_i-\e_j$ ($i<j$) and a coordinate
$\bar{x}_{i,j}$ associated with  each positive root $\e_i+\e_j$
($i<j$) respectively. These $n\times (n-1)$ coordinates satisfy
the following commutation relations: \bea
 &&[x_{i,j},x_{m,l}]=0,\,\,[\partial_{x_{i,j}},\partial_{x_{m,l}}]=0,
 \,\,[\partial_{x_{i,j}},x_{m,l}]=\d_{im}\d_{jl},\label{Fundament-Comm-1}\\
 &&[\bar{x}_{i,j},\bar{x}_{m,l}]=0,\,\,
 [\partial_{\bar{x}_{i,j}},\partial_{\bar{x}_{m,l}}]=0, \,\,
 [\partial_{\bar{x}_{i,j}},\bar{x}_{m,l}]=\d_{im}\d_{jl},
 \label{Fundament-Comm-2}
\eea and the other commutation relations are vanishing. Let
$\langle\L|$ be the highest weight vector of the representation of
$so(2n)$ with highest weights $\{\l_i\}$ , satisfying the
following conditions: \bea
 &&\langle\L|F_i=0,\qquad\qquad 1\leq i\leq n,\label{highestweight-1}\\
 &&\langle\L|H_i=\l_i\,\langle\L|,\qquad\qquad 1\leq i\leq
  n.\label{Lowestweight-2}
\eea Here the generators $H_i$ are expressed in terms of some
linear combinations of $H_{\a}$ (\ref{D-H}). An arbitrary vector
in the corresponding Verma module \footnote{The irreducible
highest weight representation can be obtained from the Verma
module through the cohomology procedure \cite{Bou90} with the help
of screening operators (e.g. (\ref{Scr-P-D-1}) and
(\ref{Scr-P-D-2}) below). } is parametrized by $\langle\L|$ and
the coordinates ($x$ and $\bar{x}$) as \bea
\langle\L,x,\bar{x}|=\langle\L|G_{+}(x,\bar{x}),\label{States-D}\eea
where $G_{+}(x,\bar{x})$ is given by (c.f. \cite{Ito90,Ras97})
\bea
   G_{+}(x,\bar{x})&=&\lt(\bar{G}_{n-1,n}\,G_{n-1,n}\rt)\,
      \lt(\bar{G}_{n-2,n-1}\bar{G}_{n-2,n}\,G_{n-2,n}G_{n-2,n-1}\rt)
      \ldots\no\\
    &&\quad \times \lt(\bar{G}_{1,2}\ldots\bar{G}_{1,n}
    \,G_{1,n}\ldots G_{1,2}\rt).
\eea Here, for $i<j$, $G_{i,j}$ and $\bar{G}_{i,j}$ are given by
\bea
  G_{i,j}=e^{x_{i,j}E_{\e_i-\e_j}},\qquad
  \bar{G}_{i,j}=e^{\bar{x}_{i,j}E_{\e_i+\e_j}}.
\eea One can define a differential operator realization
$\rho^{(d)}$ of the generators of $so(2n)$ by
\bea
 \rho^{(d)}(g)\,\langle\L,x,\bar{x}|\equiv \langle\L,x,\bar{x}|\,g,\qquad
 \forall g\in so(2n).\label{definition-D}
\eea Here $\rho^{(d)}(g)$ is a differential operator of the
coordinates $\{x_{i,j},\,\bar{x}_{i,j}\}$ associated with the
generator $g$, which can be obtained from the defining relation
(\ref{definition-D}). The defining relation also assures that  the
differential operator realization is actually a representation of
$so(2n)$. Therefore it is sufficient to give the differential
operators related to the simple roots, as the others can be
constructed through the simple ones by the commutation relations.
Using the relation (\ref{definition-D}) and the
Baker-Campbell-Hausdorff formula, after some algebraic
manipulations, we obtain the following differential operator
representation of the simple generators: \bea
  \rho^{(d)}(E_i)&=&\sum_{m=1}^{i-1}
    \lt(x_{m,i}\partial_{x_{m,i+1}}-\bar{x}_{m,i+1}\partial_{\bar{x}_{m,i}}\rt)
    +\partial_{x_{i,i+1}},\qquad 1\leq i\leq n-1,\label{Diff-D-1}\\
  \rho^{(d)}(E_n)&=&\sum_{m=1}^{n-2}
    \lt(x_{m,n-1}\partial_{\bar{x}_{m,n}}-x_{m,n}\partial_{\bar{x}_{m,n-1}}\rt)
    +\partial_{\bar{x}_{n-1,n}},\\
  \rho^{(d)}(F_i)&=&\sum_{m=1}^{i-1}\lt(x_{m,i+1}\partial_{x_{m,i}}
    -\bar{x}_{m,i}\partial_{\bar{x}_{m,i+1}}\rt)\no\\
    &&-\sum_{m=i+2}^n\lt(x_{i,m}\partial_{x_{i+1,m}}
    -x_{i,m}\bar{x}_{i,m}\partial_{\bar{x}_{i,i+1}}
    +\bar{x}_{i,m}\partial_{\bar{x}_{i+1,m}}\rt)
    -x^2_{i,i+1}\partial_{x_{i,i+1}}\no\\
    &&-x_{i,i+1}\lt[\sum_{m=i+2}^n(x_{i,m}\partial_{x_{i,m}}
    +\bar{x}_{i,m}\partial_{\bar{x}_{i,m}}
    -x_{i+1,m}\partial_{x_{i+1,m}}
    -\bar{x}_{i+1,m}\partial_{\bar{x}_{i+1,m}})\rt]\no\\
    &&+x_{i,i+1}(\l_i-\l_{i+1}),\qquad\qquad 1\leq i\leq n-1,\\
  \rho^{(d)}(F_n)&=&\sum_{m=1}^{n-2}\lt(\bar{x}_{m,n}\partial_{x_{m,n-1}}
    \hspace{-0.12truecm}-\hspace{-0.12truecm}\bar{x}_{m,n-1}\partial_{x_{m,n}}\rt)
    \hspace{-0.12truecm}-\hspace{-0.12truecm}\bar{x}^2_{n-1,n}\partial_{\bar{x}_{n-1,n}}
    +\bar{x}_{n-1,n}(\l_{n-1}\hspace{-0.12truecm}+\hspace{-0.12truecm}\l_n),\\
  \rho^{(d)}(H_i)&=&\sum_{m=1}^{i-1}\lt(x_{m,i}\partial_{x_{m,i}}
  -\bar{x}_{m,i}\partial_{\bar{x}_{m,i}}\rt)
  -\sum_{m=i+1}^{n}\lt(x_{i,m}\partial_{x_{i,m}}
  +\bar{x}_{i,m}\partial_{\bar{x}_{i,m}}\rt)+\l_i,\no\\
  &&\qquad\qquad i=1,\ldots,n.\label{Diff-D-2}
\eea

A direct computation shows that these differential operators
(\ref{Diff-D-1})-(\ref{Diff-D-2}) satisfy the $so(2n)$ commutation
relations corresponding to the simple roots and the associated
Serre relations. This implies that the differential representation
of non-simple generators can be consistently constructed from the
simple ones. Hence, we have obtained an explicit differential
realization of $so(2n)$.

\subsection{Free field realization of $so(2n)_k$}

With the help of the differential realization given by
(\ref{Diff-D-1})-(\ref{Diff-D-2}) we can construct the explicit
free field representation of the $so(2n)$ current algebra at
arbitrary level $k$ in terms of $n\times (n-1)$ bosonic $\b$-$\g$
pairs $\{(\b_{i,j},\,\g_{i,j}),\,
(\bar{\b}_{i,j}\bar{\g}_{i,j}),\, 1\leq i<j\leq n\}$  and $n$ free
scalar fields $\phi_i$, $i=1,\ldots,n$. These free fields obey the
following OPEs:\bea
&&\hspace{-0.8truecm}\b_{i,j}(z)\,\g_{m,l}(w)=-\g_{m,l}(z)\,\b_{i,j}(w)=
\frac{\d_{im}\d_{jl}}{(z-w)},\,\,1\leq i<j\leq n,\,\,1\leq
m<l\leq n,\label{OPE-F-1}\\
&&\hspace{-0.8truecm}\bar{\b}_{i,j}(z)\,\bar{\g}_{m,l}(w)=-\bar{\g}_{m,l}(z)\,
\bar{\b}_{i,j}(w)= \frac{\d_{im}\d_{jl}}{(z-w)},\,\,1\leq i<j\leq
n,\,\,1\leq m<l\leq n,\\
&&\hspace{-0.8truecm}\phi_i(z)\phi_j(w)=\d_{ij}\,
\ln(z-w),\,\,\,\,\,1\leq i,j\leq n,\label{OPE-F-2}\eea and the
other OPEs are trivial.

The free field realization of the $so(2n)$ current algebra is
obtained by the  substitution in the differential realization
(\ref{Diff-D-1})-(\ref{Diff-D-2}) of $so(2n)$, \bea
 &&x_{i,j}\longrightarrow \g_{i,j}(z),\quad \partial_{x_{i,j}}
   \longrightarrow \b_{i,j}(z),\quad 1\leq i<j\leq n,\label{sub-1}\\
 &&\bar{x}_{i,j}\longrightarrow \bar{\g}_{i,j}(z),\quad
   \partial_{\bar{x}_{i,j}} \longrightarrow \bar{\b}_{i,j}(z),
   \quad 1\leq i<j\leq n,\\
 &&\l_j\longrightarrow \sqrt{k+2(n-1)}\partial\phi_j(z)\qquad
   1\leq j\leq n.\label{sub-2}\eea
Moreover, in order that the resulting free field realization
satisfy the desirable OPE for $so(2n)$ currents, one needs to add
certain extra (anomalous) terms which are linear in
$\partial\g(z)$ and $\partial\bar{\g}(z)$  in the expressions of
the currents associated with negative roots (e.g. the last term in
the expressions of $F_i(z)$, see (\ref{Free-D-1})-(\ref{Free-D-2})
below). Here we present the results for the currents associated
with the simple roots, \bea
  E_{i}(z)&=&\sum_{m=1}^{i-1}\lt(\g_{m,i}(z)\b_{m,i+1}(z)-
    \bar{\g}_{m,i+1}(z)\bar{\b}_{m,i}(z)\rt)+\b_{i,i+1}(z),\quad
    1\leq i\leq n-1,\label{Free-D-1}\\
  E_{n}(z)&=&\sum_{m=1}^{n-2}\lt(\g_{m,n-1}(z)\bar{\b}_{m,n}(z)-
    \g_{m,n}(z)\bar{\b}_{m,n-1}(z)\rt)+\bar{\b}_{n-1,n}(z),\no\\
  F_{i}(z)&=&\sum_{m=1}^{i-1}\lt(\g_{m,i+1}(z)\b_{m,i}(z)
    -\bar{\g}_{m,i}(z)\bar{\b}_{m,i+1}(z)\rt)
    -\g^2_{i,i+1}(z)\b_{i,i+1}(z)\no\\
    &&-\sum_{m=i+2}^{n}\lt(\g_{i,m}(z)\b_{i+1,m}(z)
    -\g_{i,m}(z)\bar{\g}_{i,m}(z)\bar{\b}_{i,i+1}(z)
    +\bar{\g}_{i,m}(z)\bar{\b}_{i+1,m}(z)\rt)\no\\
    &&-\g_{i,i+1}(z)\lt[\sum_{m=i+2}^n\g_{i,m}(z)\b_{i,m}(z)
      +\bar{\g}_{i,m}(z)\bar{\b}_{i,m}(z)\rt]\no\\
    &&+\g_{i,i+1}(z)\lt[\sum_{m=i+2}^n\g_{i+1,m}(z)\b_{i+1,m}(z)+
    \bar{\g}_{i+1,m}(z)\bar{\b}_{i+1,m}(z)\rt]\no\\
    &&+\sqrt{k+2(n-1)}\g_{i,i+1}(z)\lt(\partial\phi_i(z)\hspace{-0.1truecm}
    -\hspace{-0.1truecm}\partial\phi_{i+1}(z)\rt)\no\\
    &&+(k+2(i-1))\partial\g_{i,i+1}(z),
    \quad 1\leq i\leq n-1,\no\\
  F_n(z)&=&\sum_{m=1}^{n-2}\lt(\bar{\g}_{m,n}(z)\b_{m,n-1}(z)
     -\bar{\g}_{m,n-1}(z)\b_{m,n}(z)\rt)
     -\bar{\g}^2_{n-1,n}(z)\bar{\b}_{n-1,n}(z)\no\\
     &&+\sqrt{k+2(n-1)}\bar{\g}_{n-1,n}(z)\lt(
     \partial\phi_{n-1}(z)\hspace{-0.1truecm}+\hspace{-0.1truecm}\partial\phi_{n}(z)\rt)
     \hspace{-0.1truecm}+\hspace{-0.1truecm}
     (k+2(n\hspace{-0.1truecm}-\hspace{-0.1truecm}2))\partial\bar{\g}_{n-1,n}(z),\no\\
  H_i(z)&=&\hspace{-0.16truecm}\sum_{m=1}^{i-1}\hspace{-0.16truecm}\lt(\g_{m,i}(z)\b_{m,i}(z)
     \hspace{-0.1truecm}-\hspace{-0.1truecm}\bar{\g}_{m,i}(z)\bar{\b}_{m,i}(z)\rt)
     \hspace{-0.1truecm}-\hspace{-0.1truecm}
     \sum_{m=i+1}^{n}\hspace{-0.16truecm}\lt(\g_{i,m}(z)\b_{i,m}(z)\hspace{-0.1truecm}
     +\hspace{-0.1truecm}\bar{\g}_{i,m}(z)\bar{\b}_{i,m}(z)\rt)\no\\
     &&+\sqrt{k+2(n-1)}\partial\phi_i(z),\qquad\qquad 1\leq i\leq
     n.\label{Free-D-2}
\eea Here and throughout normal ordering of free fields is implied
whenever necessary. The free field realization of the currents
associated with the non-simple roots can be obtained from the OPEs
of the simple ones. We can straightforwardly check that the above
free field realization of the currents satisfies the OPEs of the
$so(2n)$ current algebra: Direct calculation shows that there are
at most second order singularities (e.g. $1\over(z-w)^{2}$) in the
OPEs of the currents. Comparing with the definition of the current
algebra (\ref{current-OPE}), terms with first order singularity
(e.g. the coefficients of $1\over(z-w)$) are fulfilled due to the
very substitution (\ref{sub-1})-(\ref{sub-2}) and the fact that
the differential operator realizations (\ref{Diff-D-1})-
(\ref{Diff-D-2}) are a representation of the corresponding
finite-dimensional  Lie algebra $so(2n)$; terms with second order
singularity $1\over(z-w)^{2}$  also match those in the definition
(\ref{current-OPE}) after the suitable choice we made for the
anomalous terms in the expressions of the currents associated with
negative roots.

The free field realization of the $so(2n)$ current algebra
(\ref{Free-D-1})-(\ref{Free-D-2}) gives rise to  the Fock
representations of the current algebra in terms of the free fields
(\ref{OPE-F-1})-(\ref{OPE-F-2}). These representations are in
general not irreducible for the current algebra. In order to
obtain irreducible ones, one needs certain screening charges,
which are the integrals of screening currents (see
(\ref{Screen-D-1})-(\ref{Screen-D-2}) below), and performs the
cohomology procedure as in  \cite{Fat86,Fei90,Bou90,Ber89}. We
shall construct the associated screening currents in subsection 4.

\subsection{Energy-momentum tensor} \label{EMT}

In this subsection we construct the free field realization of the
Sugawara energy-momentum tensor $T(z)$ of the $so(2n)$ current
algebra. After a tedious calculation, we find \bea
  T(z)&=&\frac{1}{2\lt(k+2(n-1)\rt)}\lt\{\sum_{i<j}
     \lt(E_{\e_i-\e_j}(z)F_{\e_i-\e_j}(z)+F_{\e_i-\e_j}(z)E_{\e_i-\e_j}(z)\rt)
     \rt.\no\\
  &&+\lt.\sum_{i<j}\lt(E_{\e_i+\e_j}(z)F_{\e_i+\e_j}(z)+F_{\e_i+\e_j}(z)E_{\e_i+\e_j}(z)\rt)
     +\sum_{i=1}^nH_i(z)H_i(z)\rt\}\no\\
  &=&\sum_{i=1}^n\lt(\frac{1}{2}\partial\phi_i(z)\partial\phi_i(z)-
      \frac{n-i}{\sqrt{k+2(n-1)}}\partial^2\phi_i(z)\rt)\no\\
  &&+\sum_{i<j}\lt(\b_{i,j}(z)\partial\g_{i,j}(z)+
     \bar{\b}_{i,j}(z)\partial\bar{\g}_{i,j}(z)\rt).\label{Energy-D}
\eea It is straightforward to check that $T(z)$ satisfy the
following OPE,
\bea
   T(z)T(w)=\frac{c/2}{(z-w)^4}+\frac{2T(w)}{(z-w)^2}+\frac{\partial
        T(w)}{(z-w)}.
\eea The corresponding central charge $c$ is
\bea
 c=\frac{kn(2n-1)}{k+2(n-1)}\equiv \frac{k\,{\rm dim}(so(2n))}{k+2(n-1)}.
  \label{Center-charge-D}
\eea Moreover, we find that with regard to the energy-momentum
tensor $T(z)$ defined by (\ref{Energy-D}) the $so(2n)$ currents
associated with the simple roots (\ref{Free-D-1})-(\ref{Free-D-2})
are indeed primary fields with conformal dimensional one, namely,
\bea
  T(z)E_{i}(w)&=&\frac{E_{i}(w)}{(z-w)^2}+\frac{\partial
    E_{i}(w)}{(z-w)},\,\,1\leq i\leq n,\no\\
  T(z)F_{i}(w)&=&\frac{F_{i}(w)}{(z-w)^2}+\frac{\partial
    F_{i}(w)}{(z-w)},\,\,1\leq i\leq n,\no\\
  T(z)H_{i}(w)&=&\frac{H_{i}(w)}{(z-w)^2}+\frac{\partial
    H_{i}(w)}{(z-w)},\,\,1\leq i\leq n.\no
\eea It is expected that  the $so(2n)$ currents associated with
non-simple roots, which can be constructed through the simple
ones, are also primary fields with conformal dimensional one.
Therefore, $T(z)$ is the very energy-momentum tensor of the
$so(2n)$ current algebra.

\subsection{Screening currents} \label{SC}

Important objects in the application of free field realizations to
the computation of correlation functions  of the CFTs are
screening currents. A screening current is a primary field with
conformal dimension one and has the property that the singular
part of its OPE with the affine currents is a total derivative.
These properties ensure that the integrated screening currents
(screening charges) may be inserted into correlators while the
conformal or affine Ward identities remain intact
\cite{Dos84,Ber90}.

Free field realization of the screening currents may be
constructed from certain differential operators \cite{Bou90,Ras97}
which can be defined by the relation, \bea
 \rho^{(d)}\lt(s_{\a}\rt)\,
 \langle\L,x,\bar{x}|\equiv\langle\L|\,E_{\a}\,G_{+}(x,\bar{x}),\qquad
 {\rm for}\,\a\in\Delta_+.
 \label{Def-2}
\eea The operators $\rho^{(d)}\lt(s_{\a}\rt)$ ($\a\in\Delta_+$)
give a differential operator realization of a subalgebra of
$so(2n)$, which is spanned by $\{E_{\a},\,\a\in\Delta_+\}$. Again
it is sufficient to construct $s_i\equiv
\rho^{(d)}\lt(s_{\a_i}\rt)$ related to the simple roots. Using
(\ref{Def-2}) and the Baker-Campbell-Hausdorff formula, after some
algebraic manipulations, we obtain the following explicit
expressions for $s_i$: \bea
   s_i&=& \sum_{m=i+2}^n\lt(\bar{x}_{i+1,m}\partial_{\bar{x}_{i,m}}
     -\bar{x}_{i+1,m}x_{i+1,m}\partial_{\bar{x}_{i,i+1}}
     +x_{i+1,m}\partial_{x_{i,m}}\rt)+\partial_{x_{i,i+1}},\no\\
     &&\qquad\qquad
     1\leq i\leq n-1,\label{Scr-P-D-1}
     \\
   s_n&=&\partial_{\bar{x}_{n-1,n}}.\label{Scr-P-D-2}
\eea {}One may obtain the differential operators $s_{\a}$
associated with the non-simple generators from the above simple
ones. Following the procedure similar to \cite{Bou90,Ras97}, we
find that the free field realization of the screening currents
$S_i(z)$ corresponding to the differential operators $s_i$ is
given by \bea
 S_i(z)&=&\lt\{\sum_{m=i+2}^n\lt(\bar{\g}_{i+1,m}(z)\bar{\b}_{i,m}(z)
    -\bar{\g}_{i+1,m}(z)\g_{i+1,m}(z)\bar{\b}_{i,i+1}(z)
    +\g_{i+1,m}(z)\b_{i,m}(z)\rt)\rt.\no\\
 &&\quad+\lt.\b_{i,i+1}(z)\rt\}e^{-\frac{\a_i\cdot\vec{\phi}(z)}{\sqrt{k+2(n-1)}}},
    \quad 1\leq i\leq n-1,\label{Screen-D-1}\\
 S_n(z)&=&\bar{\b}_{n-1,n}(z)e^{-\frac{\a_{n}\cdot\vec{\phi}(z)}{\sqrt{k+2(n-1)}}}.
 \label{Screen-D-2}
\eea Here $\vec{\phi}(z)$ is \bea
 \vec{\phi}(z)=\sum_{i=1}^n\phi_i(z)\e_i.\label{Defin-Phi}
\eea The OPEs of the screening currents with the energy-momentum
tensor and the $so(2n)$ currents (\ref{Free-D-1})-(\ref{Free-D-2})
are \bea
  && T(z)S_j(w)=\frac{S_j(w)}{(z-w)^2}+\frac{\partial
       S_j(w)}{(z-w)}=\partial_w\lt\{\frac{S_j(w)}{(z-w)}\rt\},
       \,\,j=1,\ldots,n,\\
 &&E_{i}(z)S_j(w)=0,\qquad i,j=1\ldots,n,\\
 &&H_i(z)S_j(w)=0,\qquad i,j=1\ldots,n,\\
 &&F_i(z)S_j(w)=\d_{ij}\,
\partial_{w}\lt\{\frac{\lt(k+2(n-1)\rt) \,
e^{-\frac{\a_i\cdot\vec{\phi}(z)}{\sqrt{k+2(n-1)}}}}{(z-w)}\rt\},\,\,
i,j=1,\ldots,n.\eea The screening currents obtained this way are
called screening currents of the first kind \cite{Ber86}.


\section{Results for $so(2n+1)_k$}
\setcounter{equation}{0}
\subsection{Differential operator realization of $so(2n+1)$}
The root system of $B_n$ (or $so(2n+1)$) are: $\lt\{\pm\e_i\pm
\e_j|\,i\neq j,\,i,j=1,\ldots,n\rt\}$ and
$\lt\{\pm\e_i|i=1,\ldots,n\rt\}$. Among them, the positive roots
$\Delta_+$ can be chosen as: \bea
 \e_i\pm\e_j,\,{\rm for}\,1\leq i<j\leq n,\qquad {\rm and}\,\,\,\,
 \e_i,\,{\rm for}\, i=1,\ldots,n.\no
\eea The simple roots are \bea
 \a_1=\e_1-\e_2,\,\a_2=\e_2-\e_3,\ldots,
 \a_{n-1}=\e_{n-1}-\e_n,\,\a_n=\e_n.\label{simple-roots-B}
\eea Associated with each positive root $\a$, there are a raising
operator $E_{\a}$, a lowering operator $F_{\a}$ and a Cartan
generator $H_{\a}$. The matrix realization of the generators
associated with all roots of $so(2n+1)$ is given in Appendix B,
from which one may derive the structure constants for the
particular choice of the basis. Similar to the $so(2n)$ case, we
adopt the convention (\ref{convent}) for the raising/lowering
generators associated with the simple roots.

In addition to the coordinates $\{x_{i,j},\,\bar{x}_{i,j}|\,1\leq
i<j\leq n\}$, which are associated with the positive roots
$\lt\{\e_i\pm\e_j|i<j\rt\}$, we also need to introduce extra $n$
coordinates $\{x_i|i=1,\ldots,n\}$ associated with the positive
roots $\lt\{\e_i|i=1,\ldots,n\rt\}$. The coordinates
$\{x_{i,j},\,\bar{x}_{i,j}\}$ and their differentials satisfy the
same commutation relations as
(\ref{Fundament-Comm-1})-(\ref{Fundament-Comm-2}). The other
non-trivial  commutation relations are \bea
 \lt[x_i,x_j\rt]=\lt[\partial_{x_i},\partial_{x_j}\rt]=0,\qquad
 \lt[\partial_{x_i},x_j\rt]=\d_{ij}.\label{Fundament-Comm-3}
\eea  Let $\langle\L|$ be the highest weight vector of the highest
weight representation of $so(2n+1)$ satisfying the following
conditions: \bea
 &&\langle\L|F_i=0,\qquad\qquad 1\leq i\leq n,\label{highestweight-B-1}\\
 &&\langle\L|H_i=\l_i\,\langle\L|,\qquad\qquad 1\leq i\leq
  n.\label{Lowestweight-B-2}
\eea Here the generators $H_i$ are some linear combinations of
$H_{\a}$ (\ref{B-H}). An arbitrary vector in the corresponding
Verma module is parametrized by $\langle\L|$ and the coordinates
($x$ and $\bar{x}$) as \bea
\langle\L,x,\bar{x}|=\langle\L|G_{+}(x,\bar{x}),\label{States-B}\eea
where $G_{+}(x,\bar{x})$ is given by (c.f. \cite{Ito90,Ras97})
\bea
   G_{+}(x,\bar{x})&=&\lt (G_n)\,(\bar{G}_{n-1,n}\,G_{n-1}\,G_{n-1,n}\rt)\,
      \lt(\bar{G}_{n-2,n-1}\bar{G}_{n-2,n}\,G_{n-2}\,G_{n-2,n}G_{n-2,n-1}\rt)
      \ldots\no\\
    &&\quad \times \lt(\bar{G}_{1,2}\ldots\bar{G}_{1,n}\,G_1
    \,G_{1,n}\ldots G_{1,2}\rt).
\eea Here $G_{i,j}$ and $\bar{G}_{i,j}$ for $i<j$, and $G_i$ are
given by \bea
  G_{i,j}=e^{x_{i,j}E_{\e_i-\e_j}},\qquad
  \bar{G}_{i,j}=e^{\bar{x}_{i,j}E_{\e_i+\e_j}},\qquad G_i=e^{x_iE_{\e_i}}.
\eea Then one can define a differential operator realization
$\rho^{(d)}$ of the generators of $so(2n+1)$ by
\bea
 \rho^{(d)}(g)\,\langle\L,x,\bar{x}|\equiv \langle\L,x,\bar{x}|\,g,\qquad
 \forall g\in so(2n+1).\label{definition-B}
\eea After tedious calculations analogous to those in the $so(2n)$
case, we have found the differential realization of $so(2n+1)$.
Here we give the results for the generators associated with the
simple roots, \bea
  \rho^{(d)}(E_i)&=&\sum_{m=1}^{i-1}
    \lt(x_{m,i}\partial_{x_{m,i+1}}-\bar{x}_{m,i+1}\partial_{\bar{x}_{m,i}}\rt)
    +\partial_{x_{i,i+1}},\qquad 1\leq i\leq n-1,\label{Diff-B-1}\\
  \rho^{(d)}(E_n)&=&\sum_{m=1}^{n-1}
    \lt(x_{m,n}\partial_{x_{m}}-x_{m}\partial_{\bar{x}_{m,n}}\rt)
    +\partial_{x_{n}},\\
  \rho^{(d)}(F_i)&=&\sum_{m=1}^{i-1}\lt(x_{m,i+1}\partial_{x_{m,i}}
    -\bar{x}_{m,i}\partial_{\bar{x}_{m,i+1}}\rt)
    -x_i\partial_{x_{i+1}}
    +\frac{x_i^2}{2}\partial_{\bar{x}_{i,i+1}}\no\\
    &&-\sum_{m=i+2}^n\lt(x_{i,m}\partial_{x_{i+1,m}}
    -x_{i,m}\bar{x}_{i,m}\partial_{\bar{x}_{i,i+1}}
    +\bar{x}_{i,m}\partial_{\bar{x}_{i+1,m}}\rt)
    -x^2_{i,i+1}\partial_{x_{i,i+1}}\no\\
    &&-x_{i,i+1}\lt[\sum_{m=i+2}^n(x_{i,m}\partial_{x_{i,m}}
    +\bar{x}_{i,m}\partial_{\bar{x}_{i,m}}
    -x_{i+1,m}\partial_{x_{i+1,m}}
    -\bar{x}_{i+1,m}\partial_{\bar{x}_{i+1,m}})\rt]\no\\
    &&+x_{i,i+1}\lt(x_{i+1}\partial_{x_{i+1}}-x_{i}\partial_{x_{i}}
    +\l_i-\l_{i+1}\rt),\qquad\qquad 1\leq i\leq n-1,\\
  \rho^{(d)}(F_n)&=&\sum_{m=1}^{n-1}\lt(x_{m}\partial_{x_{m,n}}
    \hspace{-0.12truecm}-\hspace{-0.12truecm}\bar{x}_{m,n}\partial_{x_{m}}\rt)
    \hspace{-0.12truecm}-\hspace{-0.12truecm}\frac{x^2_{n}}{2}\partial_{x_{n}}
    +x_{n}\l_{n},\\
  \rho^{(d)}(H_i)&=&\sum_{m=1}^{i-1}\lt(x_{m,i}\partial_{x_{m,i}}
  -\bar{x}_{m,i}\partial_{\bar{x}_{m,i}}\rt)
  -\sum_{m=i+1}^{n}\lt(x_{i,m}\partial_{x_{i,m}}
  +\bar{x}_{i,m}\partial_{\bar{x}_{i,m}}\rt)\no\\
  &&-x_i\partial_{x_i}+\l_i,\qquad i=1,\ldots,n.\label{Diff-B-2}
\eea
\subsection{Free field realization of $so(2n+1)_k$}

With the help of the differential realization given by
(\ref{Diff-B-1})-(\ref{Diff-B-2}) we can construct the free field
representation of the $so(2n+1)$ current algebra with arbitrary
level $k$ in terms of $n^2$ bosonic $\b$-$\g$ pairs
$\{(\b_{i,j},\,\g_{i,j}),\, (\bar{\b}_{i,j}\bar{\g}_{i,j}),\,
1\leq i<j\leq n\}$ and $\{(\b_i,\,\g_i)|i=1,\ldots,n\}$,  and $n$
free scalar fields $\phi_i$, $i=1,\ldots,n$. The free fields
$\{(\b_{i,j},\,\g_{i,j}),\, (\bar{\b}_{i,j}\bar{\g}_{i,j})\}$ and
$\{\phi_i\}$ obey the same OPEs as
(\ref{OPE-F-1})-(\ref{OPE-F-2}). The other non-trivial  OPEs are
\bea
  \b_i(z)\g_j(w)=-\g_j(z)\b_i(w)=\frac{\d_{ij}}{(z-w)},\qquad
  i,j=1,\ldots,n.\label{OPE-F-3}
\eea

The free field realization of the $so(2n+1)$ current algebra is
obtained by the substitution  in the differential realization
(\ref{Diff-B-1})-(\ref{Diff-B-2}) of $so(2n+1)$, \bea
 &&x_{i,j}\longrightarrow \g_{i,j}(z),\quad \partial_{x_{i,j}}
   \longrightarrow \b_{i,j}(z),\quad 1\leq i<j\leq n,\no\\
 &&\bar{x}_{i,j}\longrightarrow \bar{\g}_{i,j}(z),\quad
   \partial_{\bar{x}_{i,j}} \longrightarrow \bar{\b}_{i,j}(z),
   \quad 1\leq i<j\leq n,\no\\
 &&x_i\longrightarrow \g_i(z),\qquad \partial_{x_i}\longrightarrow
  \b_i(z),\quad i=1,\ldots n,\no\\
 &&\l_i\longrightarrow \sqrt{k+2n-1}\partial\phi_i(z),\quad
   i=1,\ldots n,\no\eea
followed by an addition of anomalous terms linear in
$\partial\g(z)$ and $\partial\bar{\g}(z)$ in the expressions of
the currents. Here we present the results for  the currents
associated with the simple roots, \bea
  E_{i}(z)&=&\sum_{m=1}^{i-1}\lt(\g_{m,i}(z)\b_{m,i+1}(z)-
    \bar{\g}_{m,i+1}(z)\bar{\b}_{m,i}(z)\rt)+\b_{i,i+1}(z),\quad
    1\leq i\leq n-1,\label{Free-B-1}\\
  E_{n}(z)&=&\sum_{m=1}^{n-1}\lt(\g_{m,n}(z)\b_{m}(z)-
    \g_{m}(z)\bar{\b}_{m,n}(z)\rt)+\b_{n}(z),\no\\
  F_{i}(z)&=&\sum_{m=1}^{i-1}\lt(\g_{m,i+1}(z)\b_{m,i}(z)
    -\bar{\g}_{m,i}(z)\bar{\b}_{m,i+1}(z)\rt)
    -\g_{i}(z)\b_{i+1}(z)+\frac{1}{2}\g^2_{i}(z)
    \bar{\b}_{i,i+1}(z)\no\\
    &&-\sum_{m=i+2}^{n}\lt(\g_{i,m}(z)\b_{i+1,m}(z)
    -\g_{i,m}(z)\bar{\g}_{i,m}(z)\bar{\b}_{i,i+1}(z)
    +\bar{\g}_{i,m}(z)\bar{\b}_{i+1,m}(z)\rt)\no\\
    &&-\g_{i,i+1}(z)\lt[\sum_{m=i+2}^n\g_{i,m}(z)\b_{i,m}(z)
      +\bar{\g}_{i,m}(z)\bar{\b}_{i,m}(z)\rt]\no\\
    &&+\g_{i,i+1}(z)\lt[\sum_{m=i+2}^n\g_{i+1,m}(z)\b_{i+1,m}(z)+
    \bar{\g}_{i+1,m}(z)\bar{\b}_{i+1,m}(z)\rt]\no\\
    &&-\g^2_{i,i+1}(z)\b_{i,i+1}(z)+\g_{i,i+1}(z)\g_{i+1}(z)\b_{i+1}(z)
    -\g_{i,i+1}(z)\g_{i}(z)\b_{i}(z)\no\\
    &&+\sqrt{k+2n-1}\,\g_{i,i+1}(z)
    \lt(\partial\phi_i(z)\hspace{-0.1truecm}
    -\hspace{-0.1truecm}\partial\phi_{i+1}(z)\rt)\no\\
    &&\hspace{-0.1truecm}+\hspace{-0.1truecm}
    (k+2(i-1))\partial\g_{i,i+1}(z),
    \qquad 1\leq i\leq n-1,\no\\
  F_n(z)&=&\sum_{m=1}^{n-1}\lt(\g_{m}(z)\b_{m,n}(z)
     -\bar{\g}_{m,n}(z)\b_{m}(z)\rt)
     -\frac{1}{2}\g^2_{n}(z)\b_{n}(z)\no\\
     &&+\sqrt{k+2n-1}\,\g_{n}(z)\partial\phi_{n}(z)
     \hspace{-0.1truecm}+\hspace{-0.1truecm}
     (k+2(n-1))\partial\g_{n}(z),\no\\
  H_i(z)&=&\hspace{-0.16truecm}\sum_{m=1}^{i-1}\hspace{-0.16truecm}\lt(\g_{m,i}(z)\b_{m,i}(z)
     \hspace{-0.1truecm}-\hspace{-0.1truecm}\bar{\g}_{m,i}(z)\bar{\b}_{m,i}(z)\rt)
     \hspace{-0.1truecm}-\hspace{-0.1truecm}
     \sum_{m=i+1}^{n}\hspace{-0.16truecm}\lt(\g_{i,m}(z)\b_{i,m}(z)\hspace{-0.1truecm}
     +\hspace{-0.1truecm}\bar{\g}_{i,m}(z)\bar{\b}_{i,m}(z)\rt)\no\\
     &&-\g_i(z)\b_i(z)+\sqrt{k+2n-1}\partial\phi_i(z),\qquad\qquad 1\leq i\leq
     n.\label{Free-B-2}
\eea
\subsection{Energy-momentum tensor}

After a tedious calculation, we find that the Sugawara tensor
corresponding to the quadratic Casimir  of $so(2n+1)$ is given by
\bea
  T(z)&=&\frac{1}{2\lt(k+2n-1\rt)}\lt\{\sum_{i<j}
     \lt(E_{\e_i-\e_j}(z)F_{\e_i-\e_j}(z)+F_{\e_i-\e_j}(z)E_{\e_i-\e_j}(z)\rt)
     \rt.\no\\
  &&+\sum_{i<j}\lt(E_{\e_i+\e_j}(z)F_{\e_i+\e_j}(z)+F_{\e_i+\e_j}(z)E_{\e_i+\e_j}(z)\rt)
  \no\\
  &&+\lt.\sum_{i=1}^n\lt(E_{\e_i}(z)F_{\e_i}(z)
  +F_{\e_i}(z)E_{\e_i}(z)\rt)+\sum_{i=1}^n H_i(z)H_i(z)\rt\}\no\\
  &=&\sum_{i=1}^n\lt(\frac{1}{2}\partial\phi_i(z)\partial\phi_i(z)-
      \frac{2n-2i+1}{2\sqrt{k+2n-1}}\partial^2\phi_i(z)\rt)\no\\
  &&+\sum_{i<j}\lt(\b_{i,j}(z)\partial\g_{i,j}(z)+
     \bar{\b}_{i,j}(z)\partial\bar{\g}_{i,j}(z)\rt)
     +\sum_{i=1}^n\b_{i}(z)\partial\g_{i}(z).\label{Energy-B}
\eea It is straightforward to check that $T(z)$ satisfy the
following OPE, \bea
   T(z)T(w)=\frac{c/2}{(z-w)^4}+\frac{2T(w)}{(z-w)^2}+\frac{\partial
        T(w)}{(z-w)}.
\eea The corresponding central charge $c$ is
\bea
 c=\frac{kn(2n+1)}{k+2n-1}\equiv \frac{k\,{\rm dim}(so(2n+1))}{k+2n-1}.
  \label{Center-charge-B}
\eea Moreover, we find that with regard to the  energy-momentum
tensor $T(z)$ defined by (\ref{Energy-B}) the $so(2n+1)$ currents
associated with the simple roots (\ref{Free-B-1})-(\ref{Free-B-2})
are indeed primary fields with conformal dimensional one, namely,
\bea
  T(z)E_{i}(w)&=&\frac{E_{i}(w)}{(z-w)^2}+\frac{\partial
    E_{i}(w)}{(z-w)},\,\,1\leq i\leq n,\no\\
  T(z)F_{i}(w)&=&\frac{F_{i}(w)}{(z-w)^2}+\frac{\partial
    F_{i}(w)}{(z-w)},\,\,1\leq i\leq n,\no\\
  T(z)H_{i}(w)&=&\frac{H_{i}(w)}{(z-w)^2}+\frac{\partial
    H_{i}(w)}{(z-w)},\,\,1\leq i\leq n.\no
\eea

\subsection{Screening currents}

Free field realization of the screening currents of $so(2n+1)_k$
can be constructed from the  differential operators similar to
those of the $so(2n)_k$ case, which are defined  by the relation
\bea
 \rho^{(d)}\lt(s_{\a}\rt)\,
 \langle\L,x,\bar{x}|\equiv\langle\L|\,E_{\a}\,G_{+}(x,\bar{x}),\qquad
 {\rm for}\,\a\in\Delta_+.
 \label{Def-B-2}
\eea After some algebraic manipulations, we obtain the following
explicit expressions of $s_i$  associated with the simple roots of
$so(2n+1)$: \bea
   s_i&=& \sum_{m=i+2}^n\lt(\bar{x}_{i+1,m}\partial_{\bar{x}_{i,m}}
     -\bar{x}_{i+1,m}x_{i+1,m}\partial_{\bar{x}_{i,i+1}}
     +x_{i+1,m}\partial_{x_{i,m}}\rt)+x_{i+1}\partial_{x_i}\no\\
     &&-\frac{1}{2}x^2_{i+1}\partial_{\bar{x}_{i,i+1}}+\partial_{x_{i,i+1}},\qquad
     1\leq i\leq n-1,\\
   s_n&=&\partial_{x_{n}}.
\eea The free field realization of the screening currents $S_i(z)$
of the $so(2n+1)$ current algebra corresponding to the above
differential operators $s_i$ is given by \bea
 S_i(z)&=&\lt\{\sum_{m=i+2}^n\lt(\bar{\g}_{i+1,m}(z)\bar{\b}_{i,m}(z)
    -\bar{\g}_{i+1,m}(z)\g_{i+1,m}(z)\bar{\b}_{i,i+1}(z)
    +\g_{i+1,m}(z)\b_{i,m}(z)\rt)\rt.\no\\
   &&+\lt.\g_{i+1}(z)\b_i(z)-\frac{1}{2}\g^2_{i+1}\bar{\b}_{i,i+1}(z)+
    \b_{i,i+1}(z)\rt\}e^{-\frac{\a_i\cdot\vec{\phi}(z)}{\sqrt{k+2n-1}}},
    \quad 1\leq i\leq n-1,\label{Screen-B-1}\\
 S_n(z)&=&\b_{n}(z)e^{-\frac{\a_{n}\cdot\vec{\phi}(z)}{\sqrt{k+2n-1}}},
 \label{Screen-B-2}
\eea where $\vec{\phi}(z)$ is defined in (\ref{Defin-Phi}). From
direct calculation we find that the screening currents satisfy the
required OPEs with the energy-momentum tensor (\ref{Energy-B}) and
the $so(2n+1)$ currents (\ref{Free-B-1})-(\ref{Free-B-2}), namely,
\bea
  && T(z)S_j(w)=\frac{S_j(w)}{(z-w)^2}+\frac{\partial
       S_j(w)}{(z-w)}=\partial_w\lt\{\frac{S_j(w)}{(z-w)}\rt\},
       \,\,j=1,\ldots,n,\\
 &&E_{i}(z)S_j(w)=0,\qquad i,j=1\ldots,n,\\
 &&H_i(z)S_j(w)=0,\qquad i,j=1\ldots,n,\\
 &&F_i(z)S_j(w)=\d_{ij}\,
\partial_{w}\lt\{\frac{\lt(k+2n-1\rt) \,
e^{-\frac{\a_i\cdot\vec{\phi}(z)}{\sqrt{k+2n-1}}}}{(z-w)}\rt\},\,\,
i,j=1,\ldots,n.\eea These screening currents, (\ref{Screen-B-1})
and (\ref{Screen-B-2}),  are screening currents of the first kind
\cite{Ber86}.


\section{Results for $sp(2n)_k$}
\setcounter{equation}{0}
\subsection{Differential operator realization of $sp(2n)$}
The root system of $C_n$ (or $sp(2n)$) are: $\lt\{\pm\e_i\pm
\e_j|\,i\neq j,\,i,j=1,\ldots,n\rt\}$ and $\lt\{\pm
2\e_i|i=1,\ldots,n\rt\}$. Among them, the positive roots
$\Delta_+$ can be chosen as: \bea
 \e_i\pm\e_j,\,{\rm for}\,1\leq i<j\leq n,\qquad {\rm and}\,\,\,\,
 2\e_i,\,{\rm for}\, i=1,\ldots,n.\no
\eea The simple roots are \bea
 \a_1=\e_1-\e_2,\,\a_2=\e_2-\e_3,\ldots,
 \a_{n-1}=\e_{n-1}-\e_n,\,\a_n=2\e_n.\label{simple-roots-C}
\eea Associated with each positive root $\a$, there are a raising
operator $E_{\a}$, a lowering operator $F_{\a}$ and a Cartan
generator $H_{\a}$. The matrix realization of the generators
associated with all roots of $sp(2n)$ is given in Appendix C, from
which one may derive the structure constants for the particular
choice of the basis. Like the $so(2n)$ and $so(2n+1)$ cases, we
adopt the convention (\ref{convent}) for the raising/lowering
generators associated with the simple roots.

Similar to the $so(2n+1)$ case, besides the coordinates
$\{x_{i,j},\,\bar{x}_{i,j}|\,1\leq i<j\leq n\}$, which are
associated with the positive roots $\lt\{\e_i\pm\e_j|i<j\rt\}$, we
also need to introduce extra $n$ coordinates
$\{x_i|i=1,\ldots,n\}$ associated with the positive roots
$\lt\{2\e_i|i=1,\ldots,n\rt\}$. These coordinates
$\{x_{i,j},\,\bar{x}_{i,j}\}$ and $\{x_i\}$ satisfy the same
commutation relations as
(\ref{Fundament-Comm-1})-(\ref{Fundament-Comm-2}) and
(\ref{Fundament-Comm-3}).

Let $\langle\L|$ be the highest weight vector of the highest
weight representation of $sp(2n)$ satisfying the following
conditions: \bea
 &&\langle\L|F_i=0,\qquad\qquad 1\leq i\leq n,\label{highestweight-C-1}\\
 &&\langle\L|H_i=\l_i\,\langle\L|,\qquad\qquad 1\leq i\leq
  n.\label{Lowestweight-C-2}
\eea Here the generators $H_i$ are some linear combinations of
$H_{\a}$ (\ref{C-H}). An arbitrary vector in the corresponding
Verma module is parametrized by $\langle\L|$ and the coordinates
($x$ and $\bar{x}$) as \bea
\langle\L,x,\bar{x}|=\langle\L|G_{+}(x,\bar{x}),\label{States-C}\eea
where $G_{+}(x,\bar{x})$ is given by (c.f. \cite{Ito90,Ras97})
\bea
   G_{+}(x,\bar{x})&=&\lt (G_n)\,(\bar{G}_{n-1,n}\,G_{n-1}\,G_{n-1,n}\rt)\,
      \lt(\bar{G}_{n-2,n-1}\bar{G}_{n-2,n}\,G_{n-2}\,G_{n-2,n}G_{n-2,n-1}\rt)
      \ldots\no\\
    &&\quad \times \lt(\bar{G}_{1,2}\ldots\bar{G}_{1,n}\,G_1
    \,G_{1,n}\ldots G_{1,2}\rt).
\eea Here $G_{i,j}$ and $\bar{G}_{i,j}$ for $i<j$, and $G_i$ are
given by \bea
  G_{i,j}=e^{x_{i,j}E_{\e_i-\e_j}},\qquad
  \bar{G}_{i,j}=e^{\bar{x}_{i,j}E_{\e_i+\e_j}},\qquad G_i=e^{x_iE_{2\e_i}}.
\eea Then one can define a differential operator realization
$\rho^{(d)}$ of the generators of $sp(2n)$ by
\bea
 \rho^{(d)}(g)\,\langle\L,x,\bar{x}|\equiv \langle\L,x,\bar{x}|\,g,\qquad
 \forall g\in sp(2n).\label{definition-C}
\eea After tedious calculations analogous to those in the previous
cases, we have found the differential realization of $sp(2n)$.
Here we give the results for  the generators associated with the
simple roots, \bea
  \rho^{(d)}(E_i)&=&\sum_{m=1}^{i-1}
    \lt(x_{m,i}\partial_{x_{m,i+1}}-\bar{x}_{m,i+1}\partial_{\bar{x}_{m,i}}\rt)
    +\partial_{x_{i,i+1}},\qquad 1\leq i\leq n-1,\label{Diff-C-1}\\
  \rho^{(d)}(E_n)&=&\sum_{m=1}^{n-1}
    \lt(x_{m,n}\partial_{\bar{x}_{m,n}}+x^2_{m,n}\partial_{x_{m}}\rt)
    +\partial_{x_{n}},\\
  \rho^{(d)}(F_i)&=&\sum_{m=1}^{i-1}\lt(x_{m,i+1}\partial_{x_{m,i}}
    -\bar{x}_{m,i}\partial_{\bar{x}_{m,i+1}}\rt)
    -x_i\partial_{\bar{x}_{i,i+1}}
    -2\bar{x}_{i,i+1}\partial_{x_{i+1}}\no\\
    &&-\sum_{m=i+2}^n\lt(x_{i,m}\partial_{x_{i+1,m}}
    -x_{i,m}\bar{x}_{i,m}\partial_{\bar{x}_{i,i+1}}
    +\bar{x}_{i,m}\partial_{\bar{x}_{i+1,m}}
    +2\bar{x}_{i,m}x_{i+1,m}\partial_{x_{i+1}}\rt)\no\\
    &&-x_{i,i+1}\lt[\sum_{m=i+2}^n(x_{i,m}\partial_{x_{i,m}}
    +\bar{x}_{i,m}\partial_{\bar{x}_{i,m}}
    -x_{i+1,m}\partial_{x_{i+1,m}}
    -\bar{x}_{i+1,m}\partial_{\bar{x}_{i+1,m}})\rt]\no\\
    &&+x_{i,i+1}\lt(-x_{i,i+1}\partial_{x_{i,i+1}}
    +2x_{i+1}\partial_{x_{i+1}}-2x_{i}\partial_{x_{i}}\rt)\no\\
    &&+x_{i,i+1}\lt(\l_i-\l_{i+1}\rt),\qquad 1\leq i\leq n-1,\\
  \rho^{(d)}(F_n)&=&\sum_{m=1}^{n-1}\lt(\bar{x}^2_{m,n}\partial_{x_{m}}
    +\bar{x}_{m,n}\partial_{x_{m,n}}\rt)
    -x^2_{n}\partial_{x_{n}}
    +x_{n}\l_{n},\\
  \rho^{(d)}(H_i)&=&\sum_{m=1}^{i-1}\lt(x_{m,i}\partial_{x_{m,i}}
  -\bar{x}_{m,i}\partial_{\bar{x}_{m,i}}\rt)
  -\sum_{m=i+1}^{n}\lt(x_{i,m}\partial_{x_{i,m}}
  +\bar{x}_{i,m}\partial_{\bar{x}_{i,m}}\rt)\no\\
  &&-2x_i\partial_{x_i}+\l_i,\qquad i=1,\ldots,n.\label{Diff-C-2}
\eea
\subsection{Free field realization of $sp(2n)_k$}
With the help of the differential realization given by
(\ref{Diff-C-1})-(\ref{Diff-C-2}) we can construct the free field
representation of the $sp(2n)$ current algebra with arbitrary
level $k$ in terms of $n^2$ bosonic $\b$-$\g$ pairs
$\{(\b_{i,j},\,\g_{i,j}),\, (\bar{\b}_{i,j}\bar{\g}_{i,j}),\,
1\leq i<j\leq n\}$ and $\{(\b_i,\,\g_i)|i=1,\ldots,n\}$,  and $n$
free scalar fields $\phi_i$, $i=1,\ldots,n$. These free fields
$\{(\b_{i,j},\,\g_{i,j}),\,
(\bar{\b}_{i,j}\bar{\g}_{i,j}),\,(\b_i,\,\g_i)\}$  and
$\{\phi_i\}$ obey the same OPEs as (\ref{OPE-F-1})-(\ref{OPE-F-2})
and (\ref{OPE-F-3}).

The free field realization of the $sp(2n)$ current algebra is
obtained by the substitution in the differential realization
(\ref{Diff-C-1})-(\ref{Diff-C-2}) of $sp(2n)$, \bea
 &&x_{i,j}\longrightarrow \g_{i,j}(z),\quad \partial_{x_{i,j}}
   \longrightarrow \b_{i,j}(z),\quad 1\leq i<j\leq n,\no\\
 &&\bar{x}_{i,j}\longrightarrow \bar{\g}_{i,j}(z),\quad
   \partial_{\bar{x}_{i,j}} \longrightarrow \bar{\b}_{i,j}(z),
   \quad 1\leq i<j\leq n,\no\\
 &&x_i\longrightarrow \g_i(z),\qquad \partial_{x_i}\longrightarrow
  \b_i(z),\quad i=1,\ldots n,\no\\
 &&\l_i\longrightarrow \sqrt{k+2(n+1)}\partial\phi_i(z),\quad
   i=1,\ldots n,\no\eea
followed by an addition of anomalous terms linear in
$\partial\g(z)$ and $\partial\bar{\g}(z)$ in the expressions of
the currents. Here we present the results for  the currents
associated with the simple roots, \bea
  E_{i}(z)&=&\sum_{m=1}^{i-1}\lt(\g_{m,i}(z)\b_{m,i+1}(z)-
    \bar{\g}_{m,i+1}(z)\bar{\b}_{m,i}(z)\rt)+\b_{i,i+1}(z),\quad
    1\leq i\leq n-1,\label{Free-C-1}\\
  E_{n}(z)&=&\sum_{m=1}^{n-1}\lt(\g_{m,n}(z)\bar{\b}_{m,n}(z)+
    \g^2_{m,n}(z)\b_{m}(z)\rt)+\b_{n}(z),\no\\
  F_{i}(z)&=&\sum_{m=1}^{i-1}\lt(\g_{m,i+1}(z)\b_{m,i}(z)
    -\bar{\g}_{m,i}(z)\bar{\b}_{m,i+1}(z)\rt)
    -\g_{i}(z)\bar{\b}_{i,i+1}(z)-2\bar{\g}_{i,i+1}(z)
    \b_{i+1}(z)\no\\
    &&-\sum_{m=i+2}^{n}\lt(\g_{i,m}(z)\b_{i+1,m}(z)
    -\g_{i,m}(z)\bar{\g}_{i,m}(z)\bar{\b}_{i,i+1}(z)\rt)\no\\
    &&-\sum_{m=i+2}^{n}\lt(\bar{\g}_{i,m}(z)\bar{\b}_{i+1,m}(z)
    +2\bar{\g}_{i,m}(z)\g_{i+1,m}(z)\b_{i+1}(z)\rt)\no\\
    &&-\g_{i,i+1}(z)\sum_{m=i+2}^n\lt(\g_{i,m}(z)\b_{i,m}(z)
      +\bar{\g}_{i,m}(z)\bar{\b}_{i,m}(z)\rt)\no\\
    &&+\g_{i,i+1}(z)\sum_{m=i+2}^n\lt(\g_{i+1,m}(z)\b_{i+1,m}(z)+
    \bar{\g}_{i+1,m}(z)\bar{\b}_{i+1,m}(z)\rt)\no\\
    &&-\g^2_{i,i+1}(z)\b_{i,i+1}(z)+2\g_{i,i+1}(z)\g_{i+1}(z)\b_{i+1}(z)
    -2\g_{i,i+1}(z)\g_{i}(z)\b_{i}(z)\no\\
    &&+\sqrt{k+2(n+1)}\,\g_{i,i+1}(z)
    \lt(\partial\phi_i(z)-\partial\phi_{i+1}(z)\rt)\no\\
    &&+(k+2(i-1))\partial\g_{i,i+1}(z),
    \qquad 1\leq i\leq n-1,\no\\
  F_n(z)&=&\sum_{m=1}^{n-1}\lt(\bar{\g}^2_{m,n}(z)\b_{m}(z)
     +\bar{\g}_{m,n}(z)\b_{m,n}(z)\rt)
     -\g^2_{n}(z)\b_{n}(z)\no\\
     &&+\sqrt{k+2(n+1)}\,\g_{n}(z)\partial\phi_{n}(z)
     +(\frac{k}{2}+(n-1))\partial\g_{n}(z),\no\\
  H_i(z)&=&\hspace{-0.16truecm}\sum_{m=1}^{i-1}\hspace{-0.16truecm}\lt(\g_{m,i}(z)\b_{m,i}(z)
     \hspace{-0.1truecm}-\hspace{-0.1truecm}\bar{\g}_{m,i}(z)\bar{\b}_{m,i}(z)\rt)
     \hspace{-0.1truecm}-\hspace{-0.1truecm}
     \sum_{m=i+1}^{n}\hspace{-0.16truecm}\lt(\g_{i,m}(z)\b_{i,m}(z)\hspace{-0.1truecm}
     +\hspace{-0.1truecm}\bar{\g}_{i,m}(z)\bar{\b}_{i,m}(z)\rt)\no\\
     &&-2\g_i(z)\b_i(z)+\sqrt{k+2(n+1)}\partial\phi_i(z),\qquad\qquad 1\leq i\leq
     n.\label{Free-C-2}
\eea

\subsection{Energy-momentum tensor}

After a tedious calculation, we find that the Sugawara tensor
corresponding to the quadratic Casimir  of $sp(2n)$ is given by
\bea
  T(z)&=&\frac{1}{2\lt(k+2(n+1)\rt)}\lt\{\sum_{i<j}
     \lt(E_{\e_i-\e_j}(z)F_{\e_i-\e_j}(z)+F_{\e_i-\e_j}(z)E_{\e_i-\e_j}(z)\rt)
     \rt.\no\\
  &&+\sum_{i<j}\lt(E_{\e_i+\e_j}(z)F_{\e_i+\e_j}(z)+F_{\e_i+\e_j}(z)E_{\e_i+\e_j}(z)\rt)
  \no\\
  &&+2\lt.\sum_{i=1}^n\lt(E_{2\e_i}(z)F_{2\e_i}(z)
  +F_{2\e_i}(z)E_{2\e_i}(z)\rt)+\sum_{i=1}^n H_i(z)H_i(z)\rt\}\no\\
  &=&\sum_{i=1}^n\lt(\frac{1}{2}\partial\phi_i(z)\partial\phi_i(z)-
      \frac{n-i+1}{\sqrt{k+2(n+1)}}\partial^2\phi_i(z)\rt)\no\\
  &&+\sum_{i<j}\lt(\b_{i,j}(z)\partial\g_{i,j}(z)+
     \bar{\b}_{i,j}(z)\partial\bar{\g}_{i,j}(z)\rt)
     +\sum_{i=1}^n\b_{i}(z)\partial\g_{i}(z).\label{Energy-C}
\eea It is straightforward to check that $T(z)$ satisfy the
following OPE, \bea
   T(z)T(w)=\frac{c/2}{(z-w)^4}+\frac{2T(w)}{(z-w)^2}+\frac{\partial
        T(w)}{(z-w)}.
\eea The corresponding central charge $c$ is
\bea
 c=\frac{kn(2n+1)}{k+2(n+1)}\equiv \frac{k\,{\rm dim}(sp(2n))}{k+2(n+1)}.
  \label{Center-charge-C}
\eea Moreover, we find that with regard to the energy-momentum
tensor $T(z)$ defined by (\ref{Energy-C}) the $sp(2n)$ currents
associated with the simple roots (\ref{Free-C-1})-(\ref{Free-C-2})
are indeed primary fields with conformal dimensional one, namely,
\bea
  T(z)E_{i}(w)&=&\frac{E_{i}(w)}{(z-w)^2}+\frac{\partial
    E_{i}(w)}{(z-w)},\,\,1\leq i\leq n,\no\\
  T(z)F_{i}(w)&=&\frac{F_{i}(w)}{(z-w)^2}+\frac{\partial
    F_{i}(w)}{(z-w)},\,\,1\leq i\leq n,\no\\
  T(z)H_{i}(w)&=&\frac{H_{i}(w)}{(z-w)^2}+\frac{\partial
    H_{i}(w)}{(z-w)},\,\,1\leq i\leq n.\no
\eea

\subsection{Screening currents}

Free field realization of the screening currents of $sp(2n)_k$ can
be constructed from the  differential operators similar to the
$so(2n)_k$ and $so(2n+1)_k$ cases in previous sections, which are
defined by the relation \bea
 \rho^{(d)}\lt(s_{\a}\rt)\,
 \langle\L,x,\bar{x}|\equiv\langle\L|\,E_{\a}\,G_{+}(x,\bar{x}),\qquad
 {\rm for}\,\a\in\Delta_+.
 \label{Def-C-2}
\eea After some algebraic manipulations, we obtain the following
explicit expressions of $s_i$ associated with the simple roots of
$sp(2n)$: \bea
   s_i&=& \sum_{m=i+2}^n\lt(\bar{x}_{i+1,m}\partial_{\bar{x}_{i,m}}
     -\bar{x}_{i+1,m}x_{i+1,m}\partial_{\bar{x}_{i,i+1}}
     +x_{i+1,m}\partial_{x_{i,m}}+2x_{i+1,m}\bar{x}_{i,m}\partial_{x_i}\rt)\no\\
     &&+x_{i+1}\partial_{\bar{x}_{i,i+1}}
     +2\bar{x}_{i,i+1}\partial_{x_{i}}+\partial_{x_{i,i+1}},\qquad
     1\leq i\leq n-1,\\
   s_n&=&\partial_{x_{n}}.
\eea The free field realization of the screening currents $S_i(z)$
of the $sp(2n)$ current algebra corresponding to the above
differential operators $s_i$ is given by \bea
 S_i(z)&=&\lt\{\sum_{m=i+2}^n\lt(\bar{\g}_{i+1,m}(z)\bar{\b}_{i,m}(z)
    -\bar{\g}_{i+1,m}(z)\g_{i+1,m}(z)\bar{\b}_{i,i+1}(z)
    \rt)\rt.\no\\
   &&+\sum_{m=i+2}^n\lt(\g_{i+1,m}(z)\b_{i,m}(z)
    +2\g_{i+1,m}(z)\bar{\g}_{i,m}(z)\b_i(z)\rt)
    +\g_{i+1}(z)\bar{\b}_{i,i+1}(z)\no\\
   &&+\lt. 2\bar{\g}_{i,i+1}\b_{i}(z)+
    \b_{i,i+1}(z)\rt\}e^{-\frac{\a_i\cdot\vec{\phi}(z)}{\sqrt{k+2(n+1)}}},
    \quad 1\leq i\leq n-1,\label{Screen-C-1}\\
 S_n(z)&=&\b_{n}(z)e^{-\frac{\a_{n}\cdot\vec{\phi}(z)}{\sqrt{k+2(n+1)}}},
 \label{Screen-C-2}
\eea where $\vec{\phi}(z)$ is defined in (\ref{Defin-Phi}). From
direct calculations we find that the screening currents satisfy
the required OPEs with the energy-momentum tensor (\ref{Energy-C})
and the $sp(2n)$ currents (\ref{Free-C-1})-(\ref{Free-C-2}),
namely,  \bea
  && T(z)S_j(w)=\frac{S_j(w)}{(z-w)^2}+\frac{\partial
       S_j(w)}{(z-w)}=\partial_w\lt\{\frac{S_j(w)}{(z-w)}\rt\},
       \,\,j=1,\ldots,n,\\
 &&E_{i}(z)S_j(w)=0,\qquad i,j=1\ldots,n,\\
 &&H_i(z)S_j(w)=0,\qquad i,j=1\ldots,n,\\
 &&F_i(z)S_j(w)=\frac{\d_{ij}}{1+\d_{in}}\,
\partial_{w}\lt\{\frac{\lt(k+2(n+1)\rt) \,
e^{-\frac{\a_i\cdot\vec{\phi}(z)}{\sqrt{k+2(n+1)}}}}{(z-w)}\rt\},\,\,
i,j=1,\ldots,n.\eea These screening currents, given in
(\ref{Screen-C-1}) and (\ref{Screen-C-2}),  are screening currents
of the first kind \cite{Ber86}.


\section{Discussions}
\label{Con} \setcounter{equation}{0}

We have constructed the explicit expressions of the free field
representations for the $so(2n)$, $so(2n+1)$ and $sp(2n)$ current
algebras at an arbitrary level $k$, and the corresponding
energy-momentum tensors. We have also found the free field
representation of $n$ screening currents of the first kind for
each current algebra. Our results reduce to those in \cite{Bou90}
for the $so(5)$ case.

The free field realizations (\ref{Free-D-1})-(\ref{Free-D-2}),
(\ref{Free-B-1})-(\ref{Free-B-2}) and
(\ref{Free-C-1})-(\ref{Free-C-2}) of the current algebras
$so(2n)_k$, $so(2n+1)_k$ and $sp(2n)_k$ respectively give rise to
the Fock representations of the corresponding current algebras in
terms of the free fields (\ref{OPE-F-1}),(\ref{OPE-F-2}) and
(\ref{OPE-F-3}). They provide explicit realizations of the  vertex
operator constructions of representations for affine Lie algebras
\cite{Lep84,Pri02}. Moreover, these representations are in general
not irreducible for the current algebras. To obtain irreducible
representations, one needs the associated screening charges, which
are the integrals of the corresponding screening currents
((\ref{Screen-D-1})-(\ref{Screen-D-2}),
(\ref{Screen-B-1})-(\ref{Screen-B-2}) and
(\ref{Screen-C-1})-(\ref{Screen-C-2})), and performs the
cohomology procedures as in \cite{Fat86,Fei90,Bou90,Ber89}.

Our explicit expressions of the affine currents, energy-momentum
tensor and screening currents in terms of free fields should allow
one to construct the primary fields and correlation functions of
the associated WZNW models using the method developed in
\cite{Ras97}. The approach presented in this paper can be
generalized to construct  the explicit free field realizations of
the current superalgebra $osp(m|2n)_k$ with generic $m$ and $n$.
Results will be reported elsewhere \cite{Yan08}.

\section*{Acknowledgements}
The financial support from  the Australian Research Council is
gratefully acknowledged. WLY has also been partially supported by
the New Staff Research Grant of the University of Queensland.

\section*{Appendix A: Fundamental representation of $so(2n)$
} \setcounter{equation}{0}
\renewcommand{\theequation}{A.\arabic{equation}}

Let $e_{ij}$, $i,j=1,\ldots,n$, be an $n\times n$ matrix with
entry $1$ at the $i$th row and the $j$th column and zero
elsewhere. The  $2n$-dimensional fundamental representation of
$so(2n)$, denoted by $\rho_{0}$, is given by the following
$2n\times 2n$ matrices, \bea
 &&\hspace{-1.0truecm}\rho_{0}\lt(E_{\e_i-\e_j}\rt)=
      \lt(\begin{array}{cc}e_{ij}&0\\0&-e_{ji}\end{array}\rt),\qquad
      \rho_{0}\lt(F_{\e_i-\e_j}\rt)=
      \lt(\begin{array}{cc}e_{ji}&0\\0&-e_{ij}\end{array}\rt),
   \qquad 1\leq i<j\leq n,\\[2pt]
 &&\hspace{-1.0truecm}\rho_{0}\lt(E_{\e_i+\e_j}\rt)=
     \lt(\begin{array}{cc}0&e_{ij}-e_{ji}\\0&0\end{array}\rt),\qquad\hspace{-0.56truecm}
   \rho_{0}\lt(F_{\e_i+\e_j}\rt)=
     \lt(\begin{array}{cc}0&0\\-e_{ij}+e_{ji}&0\end{array}\rt),
   \quad 1\leq i<j\leq n,\\[2pt]
 &&\hspace{-1.0truecm}\rho_{0}\lt(H_{\e_i-\e_j}\rt)=
      \lt(\begin{array}{cc}e_{ii}-e_{jj}&0\\0&-e_{ii}+e_{jj}\end{array}\rt),
      \qquad 1\leq i<j\leq n,\\[2pt]
 &&\hspace{-1.0truecm}\rho_{0}\lt(H_{\e_i+\e_j}\rt)=
      \lt(\begin{array}{cc}e_{ii}+e_{jj}&0\\0&-e_{ii}-e_{jj}\end{array}\rt),
     \qquad 1\leq i<j\leq n.
\eea We introduce $n$ linear-independent generators $H_i$
$(i=1,\ldots n)$, \bea
 H_i=\frac{1}{2}(H_{\e_i-\e_j}+H_{\e_i+\e_j}). \label{D-H}
\eea Actually, the above generators $\{H_i\}$ span the Cartan
subalgebra of $so(2n)$. In the fundamental representation, these
generators can be realized by \bea
 \rho_{0}\lt(H_i\rt)=\lt(\begin{array}{cc}e_{ii}&0\\0&-e_{ii}\end{array}\rt),\qquad
 i=1,\ldots,n.
\eea

The corresponding nondegenerate invariant bilinear symmetric form
of $so(2n)$ is given by \bea
 (x,y)=\frac{1}{2}tr\lt(\rho_0(x)\rho_0(y)\rt),\qquad
  \forall x,y\in so(2n).\label{Bilinear-D}
\eea

\section*{Appendix B: Fundamental representation of $so(2n+1)$
} \setcounter{equation}{0}
\renewcommand{\theequation}{B.\arabic{equation}}
Let $e_i$ ($i=1,\ldots,n$) be an $n$-dimensional row vector with
the $i$th component being $1$ and all others being zero, and
$e_i^T$ be the transport of $e_i$, namely, \bea
e_i=(0,\ldots,0,1,0,\ldots,0),\qquad e_i^T=\lt(\begin{array}{l}0\\
\vdots\\0\\1\\0\\\vdots\\0\end{array}\rt).\eea Then, the
$(2n+1)$-dimensional fundamental representation of $so(2n+1)$,
denoted by $\rho_{0}$, is given by the following $(2n+1)\times
(2n+1)$ matrices, \bea
 &&\hspace{-1.0truecm}\rho_{0}\lt(E_{\e_i-\e_j}\rt)=
      \lt(\begin{array}{ccc}0&0&0\\0&e_{ij}&0\\0&0&-e_{ji}\end{array}\rt),
   \,\,\rho_{0}\lt(F_{\e_i-\e_j}\rt)=
      \lt(\begin{array}{ccc}0&0&0\\0&e_{ji}&0\\0&0&-e_{ij}\end{array}\rt),
   \quad 1\leq i<j\leq n,\\[2pt]
 &&\hspace{-1.0truecm}\rho_{0}\lt(E_{\e_i+\e_j}\rt)\hspace{-0.16truecm}=\hspace{-0.16truecm}
     \lt(\begin{array}{ccc}0&0&0\\0&0&e_{ij}-e_{ji}\\0&0&0\end{array}\rt),
     \,\,\rho_{0}\lt(F_{\e_i+\e_j}\rt)\hspace{-0.16truecm}=\hspace{-0.16truecm}
     \lt(\begin{array}{ccc}0&0&0\\0&0&0\\0&-e_{ij}+e_{ji}&0\end{array}\rt),
   \,1\leq i<j\leq n,\\[2pt]
 &&\hspace{-1.0truecm}\rho_{0}\lt(E_{\e_i}\rt)=
     \lt(\begin{array}{ccc}0&0&e_i\\-e^T_i&0&0\\0&0&0\end{array}\rt),
     \,\qquad\rho_{0}\lt(F_{\e_i}\rt)=
     \lt(\begin{array}{ccc}0&-e_i&0\\0&0&0\\e_{i}^T&0&0\end{array}\rt),
   \quad i=1,\ldots,n,\\[2pt]
 &&\hspace{-1.0truecm}\rho_{0}\lt(H_{\e_i-\e_j}\rt)=
      \lt(\begin{array}{ccc}0&0&0\\0&e_{ii}-e_{jj}&0\\0&0&-e_{ii}+e_{jj}\end{array}\rt),
      \qquad 1\leq i<j\leq n,\\[2pt]
 &&\hspace{-1.0truecm}\rho_{0}\lt(H_{\e_i+\e_j}\rt)=
      \lt(\begin{array}{ccc}0&0&0\\0&e_{ii}+e_{jj}&0\\0&0&-e_{ii}-e_{jj}\end{array}\rt),
     \qquad 1\leq i<j\leq n,\\[2pt]
 &&\hspace{-1.0truecm}\rho_{0}\lt(H_{\e_i}\rt)=
      \lt(\begin{array}{ccc}0&0&0\\0&e_{ii}&0\\0&0&-e_{ii}\end{array}\rt),
     \qquad i=1,\ldots,n.
\eea We introduce $n$ linear-independent generators $H_i$
$(i=1,\ldots n)$, \bea
 H_i=\frac{1}{2}(H_{\e_i-\e_j}+H_{\e_i+\e_j})=H_{\e_i}.\label{B-H}
\eea Actually, the above generators $\{H_i\}$ span the Cartan
subalgebra of $so(2n+1)$. Moreover, the matrix realization of
$\{H_i\}$ in the fundamental representation is given by \bea
 \rho_{0}\lt(H_i\rt)=\lt(\begin{array}{ccc}0&0&0\\0&e_{ii}&0\\0&0&-e_{ii}
 \end{array}\rt),\qquad i=1,\ldots,n.
\eea

The corresponding nondegenerate invariant bilinear symmetric form
of $so(2n+1)$ is given by \bea
 (x,y)=\frac{1}{2}tr\lt(\rho_0(x)\rho_0(y)\rt),\qquad
  \forall x,y\in so(2n+1).\label{Bilinear-B}
\eea

\section*{Appendix C: Fundamental representation of $sp(2n)$
} \setcounter{equation}{0}
\renewcommand{\theequation}{C.\arabic{equation}}

The  $2n$-dimensional fundamental representation of $sp(2n)$,
denoted by $\rho_{0}$, is given by the following $2n\times 2n$
matrices, \bea
 &&\hspace{-1.0truecm}\rho_{0}\lt(E_{\e_i-\e_j}\rt)=
      \lt(\begin{array}{cc}e_{ij}&0\\0&-e_{ji}\end{array}\rt),\qquad
      \rho_{0}\lt(F_{\e_i-\e_j}\rt)=
      \lt(\begin{array}{cc}e_{ji}&0\\0&-e_{ij}\end{array}\rt),
   \qquad 1\leq i<j\leq n,\\[2pt]
 &&\hspace{-1.0truecm}\rho_{0}\lt(E_{\e_i+\e_j}\rt)=
     \lt(\begin{array}{cc}0&e_{ij}+e_{ji}\\0&0\end{array}\rt),\quad
   \rho_{0}\lt(F_{\e_i+\e_j}\rt)=
     \lt(\begin{array}{cc}0&0\\e_{ij}+e_{ji}&0\end{array}\rt),
   \quad 1\leq i<j\leq n,\\[2pt]
 &&\hspace{-1.0truecm}\rho_{0}\lt(E_{2\e_i}\rt)=
     \lt(\begin{array}{cc}0&e_{ii}\\0&0\end{array}\rt),\qquad\qquad
   \rho_{0}\lt(F_{2\e_i}\rt)=
     \lt(\begin{array}{cc}0&0\\e_{ii}&0\end{array}\rt),
     \qquad i=1,\ldots,n,\\[2pt]
 &&\hspace{-1.0truecm}\rho_{0}\lt(H_{\e_i-\e_j}\rt)=
      \lt(\begin{array}{cc}e_{ii}-e_{jj}&0\\0&-e_{ii}+e_{jj}\end{array}\rt),
      \qquad 1\leq i<j\leq n,\\
 &&\hspace{-1.0truecm}\rho_{0}\lt(H_{\e_i+\e_j}\rt)=
      \lt(\begin{array}{cc}e_{ii}+e_{jj}&0\\0&-e_{ii}-e_{jj}\end{array}\rt),
     \qquad 1\leq i<j\leq n,\\[2pt]
 &&\hspace{-1.0truecm}\rho_{0}\lt(H_{2\e_i}\rt)=
      \lt(\begin{array}{cc}e_{ii}&0\\0&-e_{ii}\end{array}\rt),
      \qquad i=1,\ldots,n.
\eea We introduce $n$ linear-independent generators $H_i$
$(i=1,\ldots n)$, \bea
 H_i=\frac{1}{2}(H_{\e_i-\e_j}+H_{\e_i+\e_j})=H_{2\e_i}.\label{C-H}
\eea Actually, the above generators $\{H_i\}$ span the Cartan
subalgebra of $sp(2n)$. Moreover, the matrix realization of
$\{H_i\}$ in the fundamental representation is given by
\bea
 \rho_0\lt(H_i\rt)=\lt(\begin{array}{cc}e_{ii}&0\\0&-e_{ii}
 \end{array}\rt),\qquad i=1,\ldots,n.
\eea

The corresponding nondegenerate invariant bilinear symmetric form
of $sp(2n)$ is given by \bea
 (x,y)=\frac{1}{2}tr\lt(\rho_0(x)\rho_0(y)\rt),\qquad
  \forall x,y\in sp(2n).\label{Bilinear-C}
\eea



\begin{thebibliography}{99}
\bibitem{Bel84} A.\,A. Belavin, A.\,M. Polyakov and
     A.\,B. Zamolodchikov, {\it Nucl. Phys.\/} {\bf B 241} (1984), 333.
\bibitem{Fra97} P. Di Francesco, P. Mathieu and D. Senehal, {\it
     Conformal Field Theory\/}, Springer Press, Berlin, 1997.
\bibitem{Kac90} V. Kac, {\it Infinite-dimensional Lie algebras\/},
     Cambridge University Press, 1990.
\bibitem{God86} P. Goddard, A. kent and D. Olive,
     {\it Phys. Lett.\/} {\bf B 321} (1985), 88;
     {\it Commun. Math. Phys.\/} {\bf 103} (1986), 105.
\bibitem{Wak86} M. Wakimoto, {\it Commun. Math. Phys.\/}
     {\bf 104} (1986), 605.
\bibitem{Dos84} VI.\,S. Dotsenko and V.\,A. Fateev,
     {\it Nucl. Phys.\/} {\bf B 240} (1984), 312;
     {\it Nucl. Phys.\/} {\bf B 251} (1985), 3691.
\bibitem{Fat86} V.\,A. Fateev and A.\,B. Zamolodchikov,
     {\it Sov. J. Nucl. Phys.\/} {\bf 43} (1986), 657.
\bibitem{Ber90} D. Bernard and G. Felder,
     {\it Commun. Math. Phys.\/} {\bf 127} (1990), 145.
\bibitem{Fur93} P. Furlan, A.\,C. Ganchev, R. Paunov and
     V.\,B. Petkova, {\it Nucl. Phys.\/} {\bf B 394}
     (1993), 665.
\bibitem{And95} O. Andreev, {\it Phys. Lett.\/}
     {\bf B 363} (1995), 166.
\bibitem{Fei90} B. Feigin and E. Frenkel,
     {\it Commun. Math. Phys.\/} {\bf 128} (1990), 161.
\bibitem{Bou90} P. Bouwknegt, J. McCarthy and K. Pilch,
     {\it Prog. Phys. Suppl.\/} {\bf  102} (1990), 67.
\bibitem{Ber89} M. Bershadsky and H. Ooguri, {\it Phys. Lett.\/}
     {\bf B 229} (1989), 374.
\bibitem{Ito90} K. Ito and S. Komata, {\it Mod. Phys. Lett.\/}
     {\bf A 6} (1991), 581.
\bibitem{Ger90} A. Gerasimov, A. Morozov, M. Olshanetsky,
     A. Marshakov and S. Shatashvili, {\it Int. J. Mod. Phys.\/}
     {\bf A 5} (1990), 2495.
\bibitem{Fre94} E. Frenkel, {\tt hep-th/9408109}.
\bibitem{Boe97} J. de Boer and L. Feher,
     {\it Commun. Math. Phys.\/} {\bf 189} (1997), 759.
\bibitem{Ras97} J.\,L. Peterson, J. Rasmussen and M. Yu,
     {\it Nucl. Phys.\/} {\bf B 502} (1997), 649.
\bibitem{Din03} X.\,-M. Ding, M. Gould and Y.\,-Z. Zhang,
     {\it Phys. Lett.\/} {\bf A 318} (2003), 354.
\bibitem{Zha05} Y.\,-Z. Zhang, X. Liu and W.\,-L. Yang,
     {\it Nucl. Phys.\/} {\bf B 704} (2005), 510.
\bibitem{Ras98} J. Rasmussen, {\it Nucl. Phys.\/} {\bf B 510}
     (1998), 688.
\bibitem{Yan07} W.\,-L. Yang, Y.\,-Z. Zhang and X. Liu, {\it Phys.
     Lett.\/} {\bf B 641} (2006), 329; {\it J. Math. Phys.\/}
     {\bf 48} (2007), 053514.
\bibitem{Kuw90} M. Kuwahara, N. Ohta and H. Suzuki,
     {\it Nucl. Phys.\/} {\bf B 340} (1990), 448.
\bibitem{Fra96} L. Frappat, P. Sorba and A. Sciarrino,
     Dictionary on Lie algebras and superalgebras, Academic Press,
     New York, 2000.
\bibitem{Ber86} M. Bershadsky and H. Ooguri,
     {\it Commun. Math. Phys.\/} {\bf 126} (1986), 49.
\bibitem{Lep84} J. Lepowsky and R.\,L. Wilson, {\it Invent. Math.}
     {\bf 77} (1984), 199; {\it Invent. Math.} {\bf 79} (1985),
     417; J. Lepowsky and M. Primc, {\it Contemp. Math.} {\bf 46}
     (1985).
\bibitem{Pri02} M. Primc, {\tt math.QA/0205262}.
\bibitem{Yan08} W.\,-L. Yang and Y.\,-Z. Zhang,
Free field realization of the $osp(2n|2n)$ current algebra.














\end{thebibliography}
\end{document}